\documentclass{ecai}
\usepackage{times}
\usepackage{graphicx}
\usepackage{latexsym}
\usepackage{soul}
\usepackage{afterpage}

\usepackage[super]{nth}

\usepackage[utf8]{inputenc}

\usepackage[caption=false]{subfig}
\usepackage[export]{adjustbox}

\ecaisubmission   

\begin{document}

\title{Multimodal Transfer Learning-based Approaches for Retinal Vascular Segmentation}

\author{José Morano$^{1}$ \and Álvaro S. Hervella$^{1,2,}$\textsuperscript{*} \and N. Barreira$^{1,2}$ \and Jorge Novo$^{1,2}$ \and José Rouco\institute{Centro de Investigación CITIC, Universidade da Coruña, A Coruña, Spain, email: j.morano@udc.es}\textsuperscript{\ \ $,$}\institute{VARPA Research Group, Instituto de Investigación Biomédica de A Coruña (INIBIC), Universidade da Coruña, A Coruña, Spain, email: \{a.suarezh, noelia.barreira, jnovo, jrouco\}@udc.es \newline * Corresponding author, email: a.suarezh@udc.es}
}

\maketitle
\bibliographystyle{ecai}

\begin{abstract}
  In ophthalmology, the study of the retinal microcirculation is a key issue in the analysis of many ocular and systemic diseases, like hypertension or diabetes. This motivates the research on improving the retinal vasculature segmentation. Nowadays, Fully Convolutional Neural Networks (FCNs) usually represent the most successful approach to image segmentation. However, the success of these models is conditioned by an adequate selection and adaptation of the architectures and techniques used, as well as the availability of a large amount of annotated data. These two issues become specially relevant when applying FCNs to medical image segmentation as, first, the existent models are usually adjusted from broad domain applications over photographic images, and second, the amount of annotated data is usually scarcer.

  In this work, we present multimodal transfer learning-based approaches for retinal vascular segmentation, performing a comparative study of recent FCN architectures. In particular, to overcome the annotated data scarcity, we propose the novel application of self-supervised network pretraining that takes advantage of existent unlabelled multimodal data. The results demonstrate that the self-supervised pretrained networks obtain significantly better vascular masks with less training in the target task, independently of the network architecture, and that some FCN architecture advances motivated for broad domain applications do not translate into significant improvements over the vasculature segmentation task.
\end{abstract}


\section{INTRODUCTION}

The retinal vascular analysis plays a fundamental role in the diagnosis of many relevant diseases. Some of these diseases are either opthalmic or systemic, and almost all of them are of great relevance due to its incidence (\textit{e.g.} age-related macular degeneration ---AMD--- and diabetic retinopathy \cite{Abramoff:2010:Retinal_Imaging}). In this type of analysis, a common preliminary step is vascular segmentation. This step allows the measurement of different blood vessel features (such as length, width or tortuosity) that offer a proven relevance in evaluating and monitoring some frequent diseases, like diabetic retinopathy. However, the manual vascular segmentation process is a repetitive, time-consuming and challenging task, and it is very common that the vasculature segmentation masks provided by different experts present non-trivial differences, specially in microvasculature \cite{Hoover:STARE:2000}.

Different approaches have been proposed for the automation of the vascular segmentation. Initially, existing approaches were mainly based on ad-hoc image processing techniques such as vessel tracking \cite{Tolias:Fuzzy:1998}, adaptive local thresholding \cite{Jiang:Adaptive:2003} or deformable models \cite{Nain:Shape:2004}. Classical supervised learning methods, such as k-nearest neighbors \cite{Staal:VS_k-NN:2004} or Artificial Neural Networks (ANNs) \cite{Sinthanayothin:Localisation:1999, Marin:SupervisedVS:2011} have been also applied. Over the recent years, many approaches use deep learning and, more specifically, Convolutional Neural Networks (CNNs) \cite{Liskowski:VesselSegmentationDNN:2016}. This type of ANNs allows to learn features during the training, as opposed to classical approaches, where the feature extraction stages were manually designed. This allows training the networks end-to-end, in a simpler and more effective way. In the case of segmentation tasks, Fully Convolutional Neural Networks (FCNs), first presented in \cite{Long:2015:FullyConv}, meant an important advance \cite{Ronneberger:2015:U-Net}. These CNNs are built with components that only operate in local regions of the input and only depend on relative spatial coordinates. Thus, they can deal with arbitrary image sizes in a direct way, without needing to iterate over images using fixed-size patches. Nowadays, the state-of-the-art approaches in retinal vasculature analysis frequently use FCNs. Some examples are vasculature segmentation \cite{Dasgupta:VesselSegmentationFCNN:2017, Fu:VSFully:2016, Jin:DUNet:2019} or vascular feature detection \cite{Hervella:RVDetection:2019}.

The application of supervised deep learning approaches, including FCNs, presents two main drawbacks. First, the selection and adaptation of the network architectures and learning techniques is necessary to solve a specific problem. This is mitigated by the fact that the techniques that are used to solve similar problems are usually appropriate. However, most of the recent deep learning advances in image analysis are motivated by broad domain applications, that raise different challenges than those faced by medical image analysis applications. Second, the most successful solutions are usually based on supervised learning of huge networks, which require a large amount of annotated data. In medical imaging, this is specially relevant, since the image dataset annotation task is usually difficult, time-consuming and requires the participation of expert clinicians.

Several techniques have been proposed to mitigate the effect of scarce labelled data, including data augmentation \cite{Krizhevsky:AlexNet:ImageNetDCNN}, transfer learning \cite{Girshick:Hierarchies:2014} and self-supervised learning \cite{Raina:Self-taught:2007}. Data augmentation is a commonly used technique that consists in increasing the amount and diversity of training images through random transformations and variations of the original images in the training set \cite{Krizhevsky:AlexNet:ImageNetDCNN}. Another widely used technique is transfer learning, which is based on the idea of learning reuse. An usual way to implement transfer learning is to pretrain an ANN for one task and reuse the learned parameters (or a part of them) to initialize the ANN in the training process for the target task. It has been demonstrated that transfer learning can significantly improve the network performance in image-related applications \cite{Oquab:TransferLearningCNN:2014}. An usual source task for transfer learning is broad domain image classification using the ImageNet large scale dataset \cite{Deng:2009:ImageNet}. This pretraining gives rise to general-enough image features to successfully apply transfer learning in medical imaging applications \cite{Shin:Computer-Aided:2016}. However, the reuse of ImageNet pretraining is more straightforward for image classification target tasks than, for example, image segmentation. In this sense, the reuse of a pretrained ImageNet classification backend for image segmentation usually conditions the resulting network architecture \cite{Chen:DeepLab:2018, Zhao:PSPNet:2017}.

Nowadays, it is also common to use self-supervised learning tasks as pretraining. In self-supervised learning, surrogate supervised learning tasks in which the target labels can be automatically obtained from unlabelled datasets are proposed with the objective of extracting relevant patterns and features. These tasks can consist in several prediction problems, such as colorization \cite{Cheng:Colorization:2015}, or the prediction of image patches from their surrounding context \cite{Pathak:CE_Inpainting:2016}. The advantage of this approach is that pretraining can be performed in the same image domain, using the available unlabelled images. Additionally, it provides a wider source of pretraining tasks, allowing to design and adjust the network architectures according to the target tasks, while pretraining them with similar source tasks.

In medical imaging, a large amount of unlabelled images is usually available from the clinical practice routine. Even more, it is common to have several image modalities available for the same patient. This motivated the proposal of the multimodal prediction as a candidate self-supervised pretraining task between ophthalmological modalities \cite{Hervella:Understanding:2018}. Specifically, it has been proposed to use a FCN to predict fluorescein angiographies (that specially highlight the vascular tree) from retinographies, demonstrating that unsupervised understanding of relevant retinal patterns, like the retinal vasculature, emerges from this training. However, the potential of this approach as pretraining task for retinal vascular segmentation has not been explored yet.

In this work, we propose multimodal transfer learning-based approaches for retinal vascular segmentation using the multimodal retinography-angiography prediction pretraining of FCNs. With this issue, we adapted and performed a comparative study of the performance of different representative modern fully convolutional architectures (U-Net \cite{Ronneberger:2015:U-Net}, FC-DenseNet \cite{Jegou:FC-DenseNet:2017} and ENet \cite{Paszke:ENet:2017, Canziani:ENet-Embedded:2017}), being also compared to their corresponding baseline without pretraining.


\section{METHODOLOGY}

\begin{figure}[tb!]
  \centerline{\includegraphics[width=0.47\textwidth]{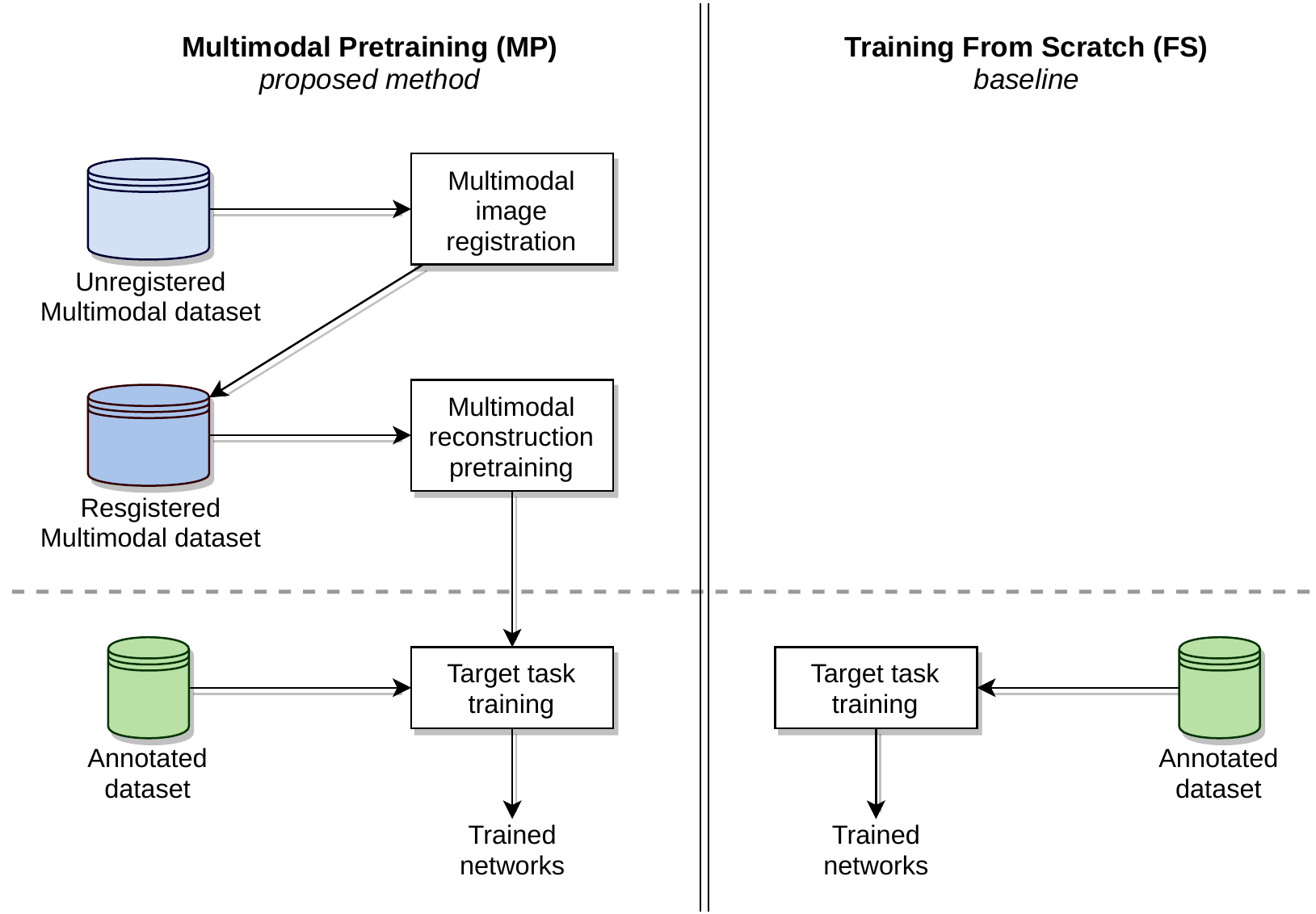}}
  \caption{Proposed Multimodal Pretraining method vs Training From Scratch baseline.}
  \label{fig:approaches}
\end{figure}

\begin{figure}[tb]
  \centerline {\includegraphics[width=0.43\textwidth]{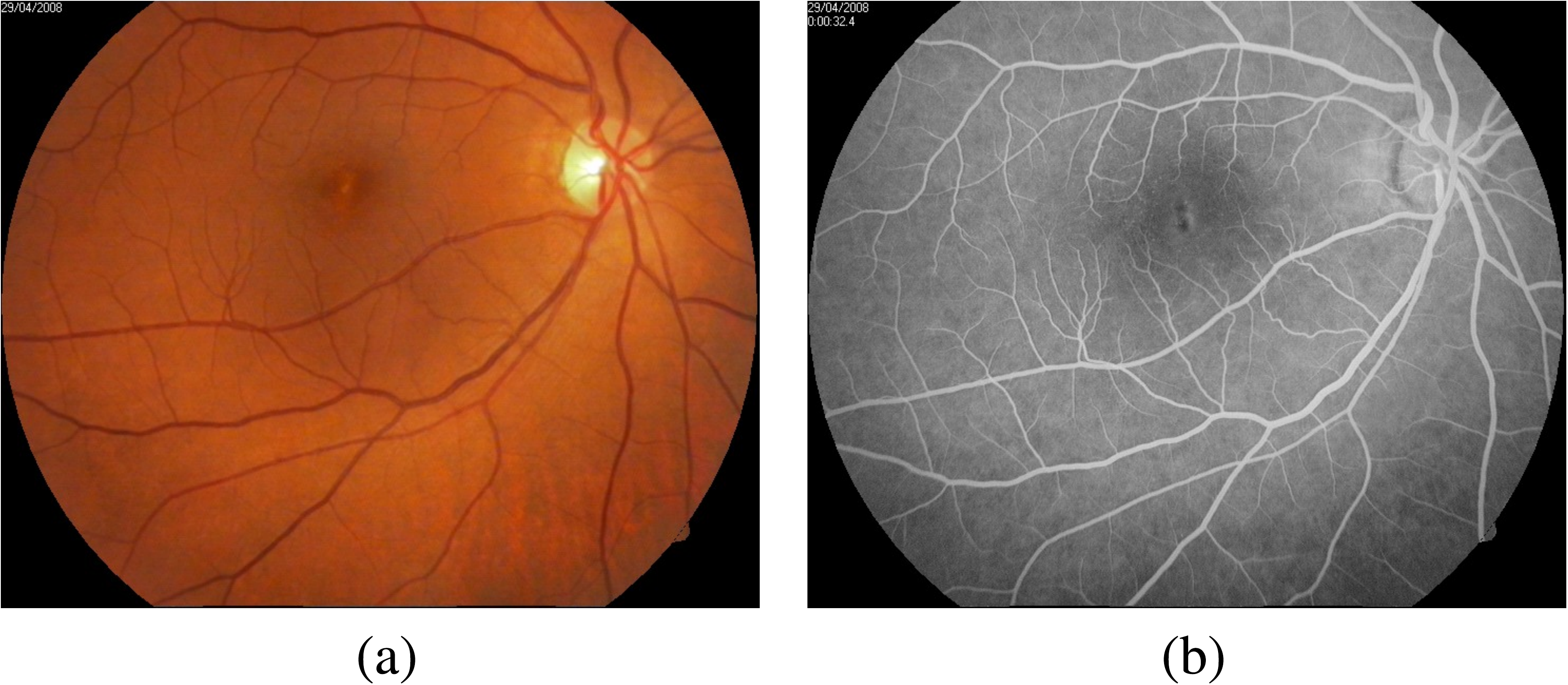}}
  \caption{Example retinography-angiography pair. (a) Retinography. \mbox{(b) Angiography}.}
  \label{fig:multimodal_pair}
\end{figure}

This work is focused on the segmentation of retinal vessel trees from retinographies using FCNs. The proposed training strategy, Multimodal Pretraining (MP), depicted in Figure \ref{fig:approaches},
is based on the pretraining of the networks on multimodal image predictions. Following this strategy, from a given unlabelled multimodal dataset that is composed of retinography-angiography image pairs for each patient (see Figure \ref{fig:multimodal_pair}), 
the FCNs are first pretrained to predict the angiography of a patient from the corresponding retinography of the same patient. To do that, the image pairs are first registered so that the pixel-wise association between the source and target modalities is available during the training. Finally, the FCNs are refined using supervised learning on a labelled dataset for the target vascular segmentation task.

In order to evaluate the proposed approach, several FCN architectures are evaluated, in comparison with the same networks trained from scratch (FS), \textit{i.e.} using a random initialization method. The following sections provide a detailed explanation of the different stages of the proposed approaches, the FCN architectures that were adapted and studied, and the corresponding experimentation details.


\subsection{Multimodal registration}

The retinography-angiography multimodal registration method proposed in \cite{Hervella:registration:2018} is used to perform the registration of the paired dataset. This method combines feature-based and intensity-based alignment approaches. In particular, the former uses domain-specific landmarks, such as vessel crossovers and bifurcations to perform an initial broad alignment. The latter employs a domain-adapted similarity metric constructed by combining a vessel enhancement preprocessing with the normalized cross-correlation similarity metric to guide a rigid and non-rigid transform optimization process \cite{Hervella:registration:2018}. Generally, this second stage allows the refinement of the initial alignment.


\subsection{Multimodal reconstruction pretraining}

The multimodal reconstruction task (MR) consists in obtaining an image with the appearance of an angiography (pseudoangiography) from its corresponding retinography. As represented in Figure \ref{fig:approaches}, trained models obtained from this task are used as pretraining models for retinal vascular segmentation in the proposed multimodal pretraining approach. To that end, a FCN is trained with registered retinography-angiography pairs.

Inspired by \cite{Hervella:Understanding:2018}, the similarity between pseudoangiographies and angiographies is measured using Structural Similarity Index (SSIM) \cite{Wang:SSIM:2004}. SSIM is usually obtained by integrating local similarity calculations along the image. Formally, a \textbf{SSIM} map is obtained as

\begin{equation}
  \textbf{SSIM}(\textbf{x},\textbf{y})=\frac{\left(2\mu_{\textbf{x}}\mu_{\textbf{y}}+C_1\right)+\left(2\sigma_{\textbf{xy}}+C_2\right)}{\left(\mu_{\textbf{x}}^2+\mu_{\textbf{y}}^2+C_1\right)+\left(\sigma_{\textbf{x}}^2+\sigma_{\textbf{y}}^2+C_2\right)}\ ,
\end{equation}

where \textbf{x} and \textbf{y} denote two single channel images, $\mu_{\textbf{x}}$ and $\mu_{\textbf{y}}$ the local averages of $\textbf{x}$ and $\textbf{y}$, respectively, $\sigma_{\textbf{x}}$ and $\sigma_{\textbf{y}}$ the local standard deviations of \textbf{x} and \textbf{y}, respectively, $\sigma_{\textbf{xy}}$ the local covariance between $\textbf{x}$ and $\textbf{y}$ and $C_1$ and $C_2$ are constants used to avoid the instability when denominator terms are close to 0. The local statistics are computed using Gaussian windows of $\sigma = 1.5$, centered at each image positions. As its output values are between 0 and 1 (being 1 a perfect similarity), the loss function is defined as

\begin{equation}
  \mathcal{L}_{SSIM}\left(\textbf{f}(\textbf{r}),\textbf{a}\right) = -\sum_{\Omega} \textbf{SSIM}\left(\textbf{f}(\textbf{r}),\textbf{a}\right)\ ,
\end{equation}

where $\textbf{f}(\textbf{r})$ denotes the network output for a given retinography \textbf{r}, \textbf{a} the target angiography and $\Omega$ the multimodal Region Of Interest (ROI). This multimodal ROI $\Omega$ is obtained by performing an \textsc{and} operation between the angiography and retinography ROIs of each given pair.


\subsection{Retinal vascular segmentation}

The retinal vascular segmentation task consists in obtaining a vascular segmentation mask with the retinal vessel tree separated from the background from a given retinography image. In order to train a FCN to perform this task, a manually labelled dataset with ground truth segmentations for each image is used.

As segmentation loss, we use the Binary Cross-Entropy (BCE) between the manually annotated vascular segmentations masks and the network output. This loss function is defined as

\begin{equation}
  \mathcal{L}_{BCE}\left(\textbf{f}(\textbf{r}),\textbf{s}\right) = -\sum\left[\textbf{s}\ log\left(\textbf{f}(\textbf{r})\right)+\left(1-\textbf{s}\right)\ log\left(1-\textbf{f}(\textbf{r})\right)\right]\ ,
\end{equation}

where $\textbf{f}(\textbf{r})$ is the network output and $\textbf{s}$ is the manually annotated vascular segmentation mask.


\subsection{Network architectures}

In order to explore the potential and capabilities of the multimodal transfer learning-based methodology herein presented, we adapted different representative modern fully convolutional architectures: \mbox{U-Net} \cite{Ronneberger:2015:U-Net}, Fully-Convolutional DenseNet \cite{Jegou:FC-DenseNet:2017} and Efficient Neural Network \cite{Paszke:ENet:2017, Canziani:ENet-Embedded:2017}.


\subsubsection{U-Net}

U-Net is a FCN architecture that was originally proposed for biomedical image segmentation \cite{Ronneberger:2015:U-Net}. It is usually considered a reference segmentation network architecture in computer vision, showing a great performance in multiple relevant tasks. It was successfully tested for multimodal reconstruction \cite{Hervella:Understanding:2018}, so it is adopted in this work as baseline model of reference.

\begin{figure}[tbp]
  \centerline{\includegraphics[width=0.47\textwidth]{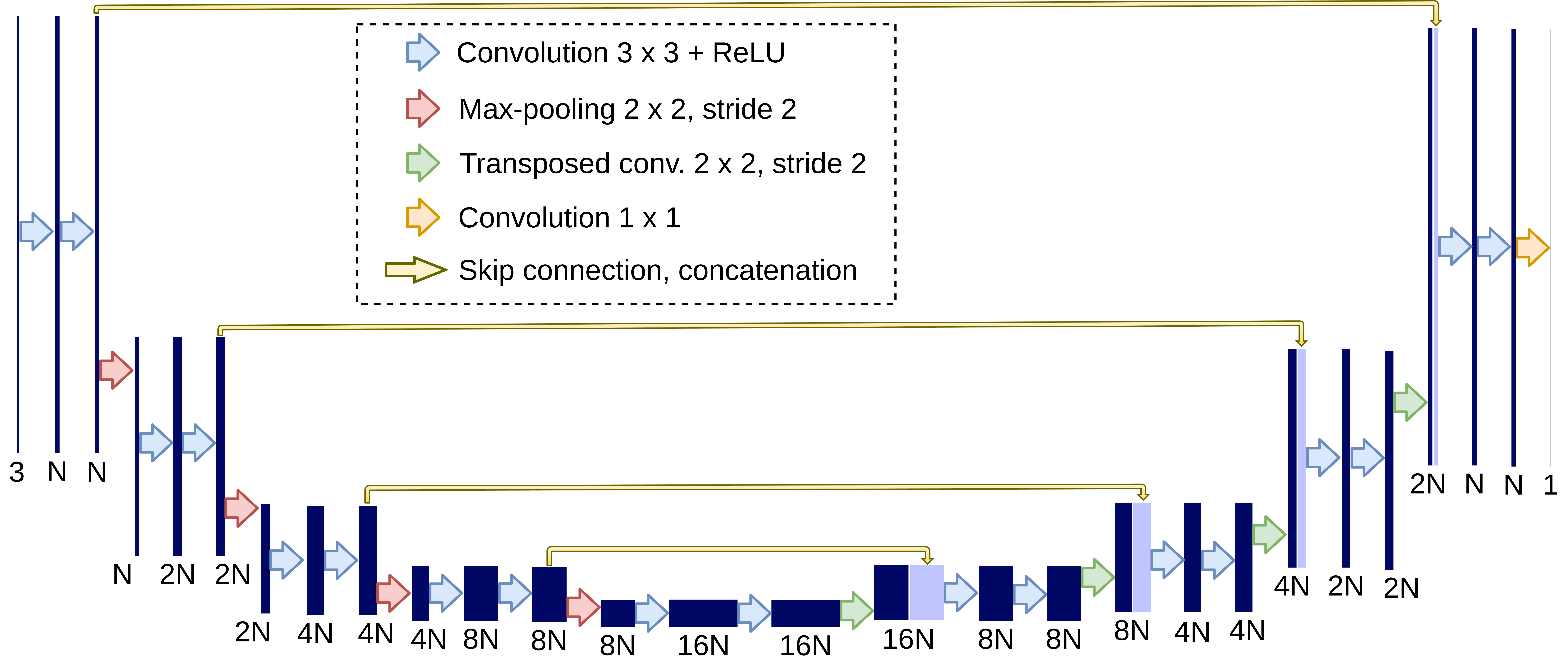}}
  \caption{U-Net architecture. $N$ corresponds to the number of base channels, that in our case is $N=64$.}
  \label{fig:U-Net}
\end{figure}

Figure \ref{fig:U-Net} shows an scheme of the used U-Net,
as proposed in \cite{Hervella:Understanding:2018}. The U-Net is characterized by a symmetric encoder-decoder architecture with skip connections between the encoder and the decoder blocks. The skip connections use transposed convolutions and concatenations to combine the channels.

The contracting path (encoder) consists of multiple subsampling blocks using convolutional layers and max-pooling. In each subsampling step, the number of channels is duplicated, and the image resolution is halved, so that the model is forced to learn high level representations through a spatial bottleneck. The expanding path (decoder) is composed of multiple upsampling blocks, in a nearly symmetrical way with respect to the contracting path. The upsampling blocks use transposed convolutions to recover the image resolution. The result is then concatenated with the corresponding feature maps from the contracting path through the skip connections, to finally apply the convolutional layers of the block.


\subsubsection{Fully Convolutional DenseNet}

\begin{figure}[b]
  \centerline{\includegraphics[width=0.47\textwidth]{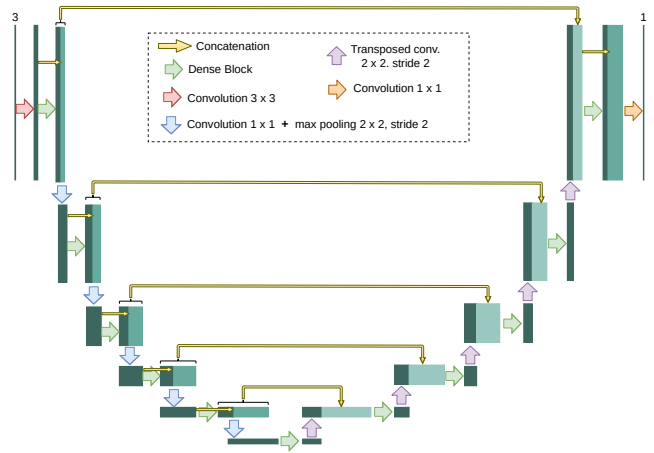}}
  \caption{FC-DenseNet architecture.}
  \label{fig:FCDN}
\end{figure}

Fully Convolutional DenseNet (FC-DenseNet) \cite{Jegou:FC-DenseNet:2017} is another FCN that uses a similar approach to U-Net, but also incorporating some ideas from DenseNet \cite{Huang:2017:DenseNet}. It was originally proposed for semantic segmentation of street images, obtaining state-of-the-art results in CamVid \cite{Brostow:CamVid:2008} and Gatech \cite{Raza:Gatech:2013} datasets. In Figure \ref{fig:FCDN},
an overview of the FC-DenseNet architecture is illustrated. Like U-Net, it follows an encoder-decoder architecture with skip connections between them and a nearly symmetrical design. However, in contrast with the U-Net model, dense blocks are used instead of the regular convolutions. These blocks were originally proposed for DenseNet, and they consist of a number of convolutional layers that are directly connected to all the subsequent layers inside the block. Each of the layers has a constant number of additional features (growth rate) that is concatenated to the outputs of all the previous layers inside the blocks. Thus, they allow the creation of complex features with less additional neurons per layer. In addition to the dense blocks, an $1 \times 1$ convolution is added before the max-pooling operation to perform feature selection and reduce the number of parameters of the subsequent blocks.

In the original proposal of the FC-DenseNet, three specific models were presented using 56, 67 and 103 convolutional layers, respectively. The only differences between them are the growth rate and the number of layers per dense block. We include all these three networks in the comparison.

The design of the FC-DenseNet allows to create much deeper models than U-Net with three times less parameters. This characteristic could be an advantage over U-Net, since it has been noticed that adding more layers usually allows the networks to create more complex representations of the data, producing a better generalization with less parameters \cite{Szegedy:googlenet:2015, Huang:DeepResNet:2016}.


\subsubsection{Efficient Neural Network}

The third adopted model is the Efficient Neural Network (ENet) architecture \cite{Paszke:ENet:2017, Canziani:ENet-Embedded:2017}. This architecture was originally proposed for semantic segmentation on street images, like FC-DenseNet. The main objective of the architecture design is to provide an efficient model (less computationally demanding) with an adequate performance in semantic segmentation tasks.

\begin{figure*}
  \centerline{\includegraphics[width=0.9\textwidth]{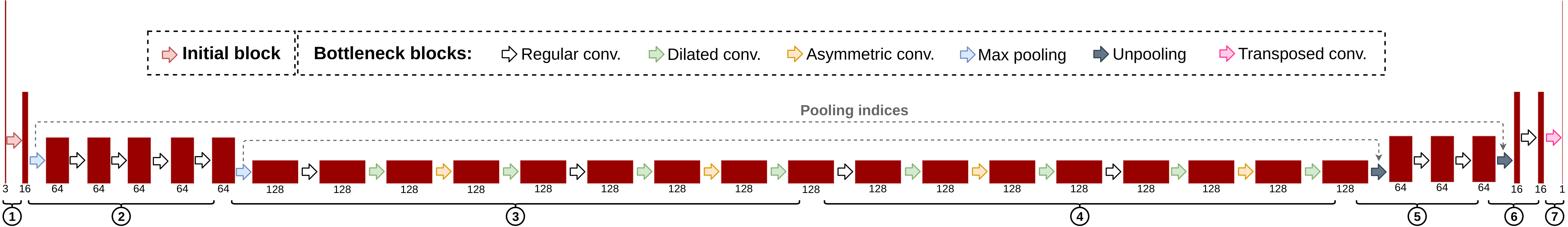}}
  \caption{ENet architecture. The circled numbers indicate the different stages of the network.}
  \label{fig:ENet}
\end{figure*}

The network architecture, depicted in Figure \ref{fig:ENet},
is organized in seven stages. As with the other two FCNs in this work, the network is composed of a contracting path (first three stages) and an expanding path (last three stages). To preserve the spatial information, the network uses skip connections to pass the pooling indices from each max pooling operation in the contracting path to the corresponding unpooling in the expanding path. Additionally, the image resolution is lowered to an smaller extent, and the network includes dilated convolutions to increase the effective receptive fields at the bottleneck.

The initial stage reduces the image resolution on each dimension through both convolutional layer and max-pooling operation, while the last 2 stages recover the original resolution to provide the output. The other stages are composed of residual blocks, following an strategy similar to ResNet \cite{He:2016:resnet}, that add the output of the convolutional layers to the block input to allow a more effective training of deeper networks. Additionally, these blocks use $1 \times 1$ convolutions to reduce the number of features before and after the spatial convolution operations. Apart from this, the network design is heterogeneous, characterized by the presence of dilated, asymmetric and transposed convolutions depending on the specific block.

This network is selected because it incorporates an alternative resolution preserving strategy compared with U-Net and FC-DenseNet (\textit{i.e.} unpooling and dilated convolutions), and because it has much less parameters than these architectures with a very deep design. In this way, by evaluating this network, we can see if lightweight deep FCNs like this are capable of performing tasks as challenging as is the retinal vascular segmentation.


\subsection{Datasets}

To perform the multimodal pretraining, we used the publicly available Isfahan MISP dataset \cite{Kashefpur:Isfahan:2016}. This dataset consists of 59 retinography-angiography unregistered image pairs from both healthy (29) and pathological (30) cases. The pathological retinographies are from patients with diabetic retinopathy. All the images present a resolution of $720 \times 576$ pixels, with a trimmed circular Region Of Interest (ROI). The pretraining of the networks is performed with all the retinograhpy-angiography pairs randomly divided into a training and validation sets of 44 and 15 pairs, respectively.

The publicly available DRIVE dataset \cite{Staal:VS_k-NN:2004} consists of 40 retinographies divided into training and test subsets (DRIVE-train and DRIVE-test), each one with 20 images. 7 of these images are from patients with diabetic retinopathy. The image sizes are $768 \times 584$, with a complete circular ROI. For all the images, a manual segmentation of the retinal vessels is provided. For training the networks, we use the 20 images from the training subset, leaving the rest for testing. To measure the robustness of the methodology with few training examples, trainings with different number of images were performed. To build the different sets, 15 images from DRIVE-train were randomly selected. From these images, we built training sets of 1, 5, 10, and 15 images, including all the images from a set in all the larger sets. In each case, the rest of the 20 images that were not included in the training set are used as validation set. Notice that training with limited annotated data is viable due to the segmentation loss providing trainining feedback for every pixel. This results in a relatively much more feedback than using image-wise labels.

The publicly available STARE dataset \cite{Hoover:STARE:2000} contains 397 retinographies from both healthy and pathological patients. Image sizes are $700 \times 605$, with a trimmed circular ROI. Retinal vascular segmentation ground truth is provided for 20 of these images. This latter images were used along the images from DRIVE-test to evaluate the retinal vascular segmentation task.


\subsection{Training details}

The optimization algorithm used to train the models is Adam \cite{Kingma:Adam:2015}. The parameters of the algorithm were empirically set as follows: the initial learning rate is set to $\alpha = 1 \times 10^{-4}$, and $\beta_1$ and $\beta_2$ are set to $0.9$ and $0.999$, respectively. The learning rate is reduced by a factor of $10$ when the validation loss does not improve for 25 epochs. In the multimodal pretraining, early stopping is applied when the validation loss does not decrease for 100 epochs. In the retinal vascular segmentation training, early stopping is applied when the validation loss does not improve for a period in which 3900 images are presented to the network. This latter early stopping, depending on the number of images instead of epochs, is used to allow the comparison between the networks using different training set sizes.

In the cases where transfer learning was not applied, the parameters of the networks were initialized following the He \textit{et al.} method with uniform distribution \cite{He:2015:Initialization}.

To artificially increase the size of the datasets, we used online data augmentation through random transformations during each epoch. These transformations consist of slight affine transformations, color and intensity variations, and vertical/horizontal flipping.


\section{RESULTS AND DISCUSSION}

In order to evaluate the performance of the presented segmentation approaches, Precision-Recall (PR) and Receiver Operating Characteristic (ROC) analysis were measured. Also, to summarize the ROC and PR curves, we calculated the Area Under the Curve (AUC) of each case. Given the clear unbalance between the vascular regions and the background eye fundus, we selected AUC-PR as the primary measurement to test and compare the performance of the networks.

\begin{figure*}
  \centerline {
  \includegraphics[width=0.33\textwidth, valign=t]{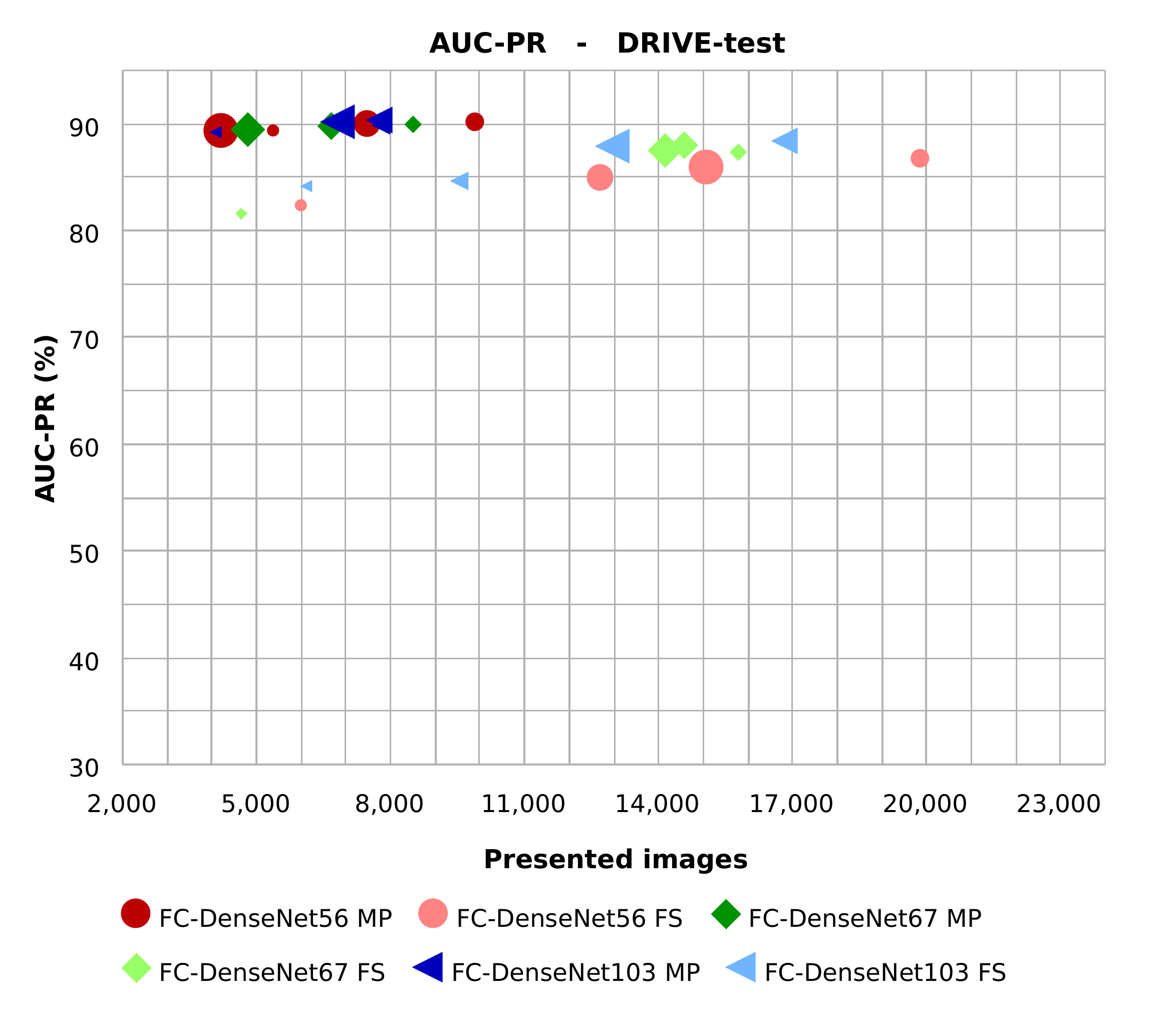}
  \includegraphics[width=0.33\textwidth, valign=t]{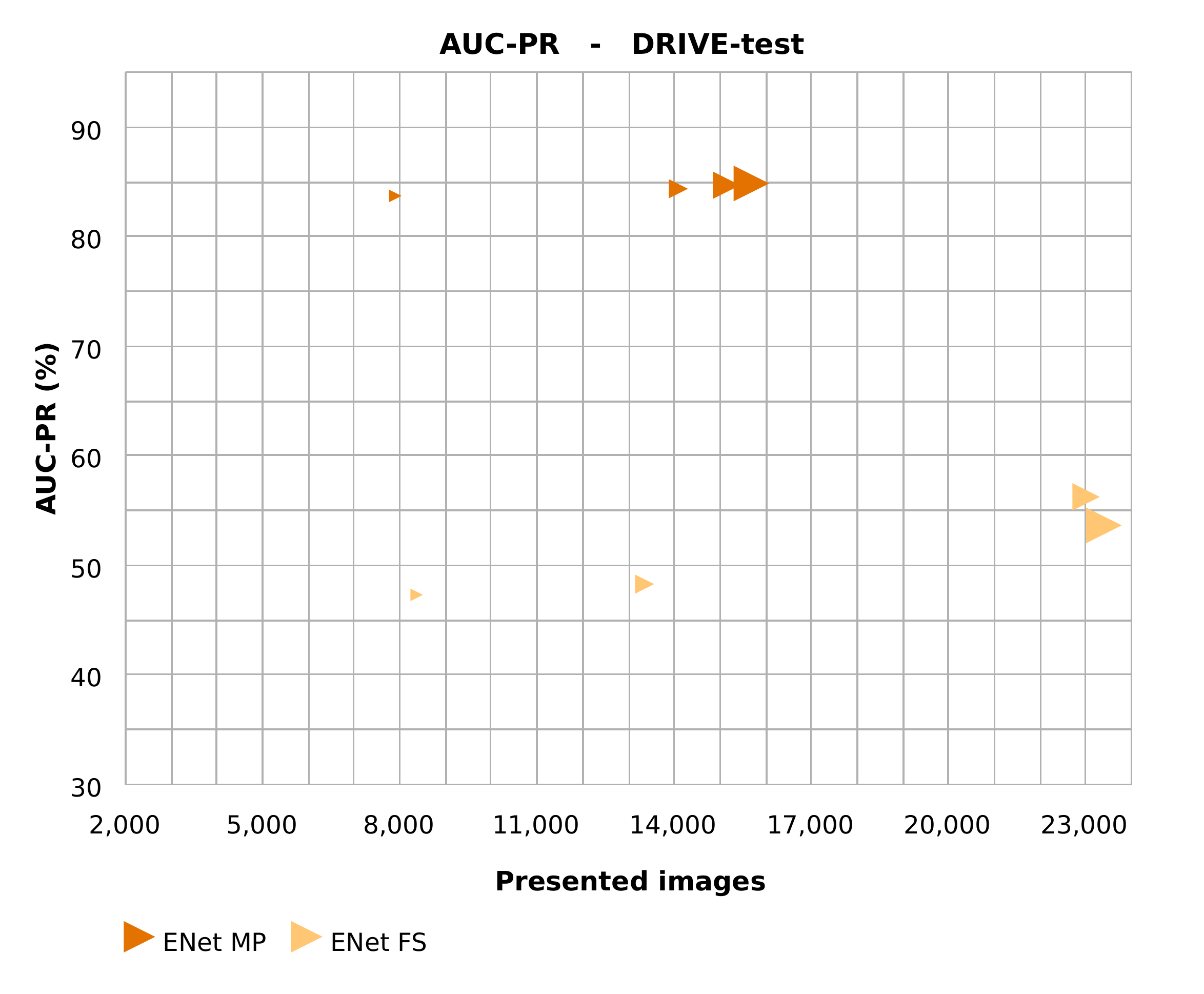}
  \includegraphics[width=0.33\textwidth, valign=t]{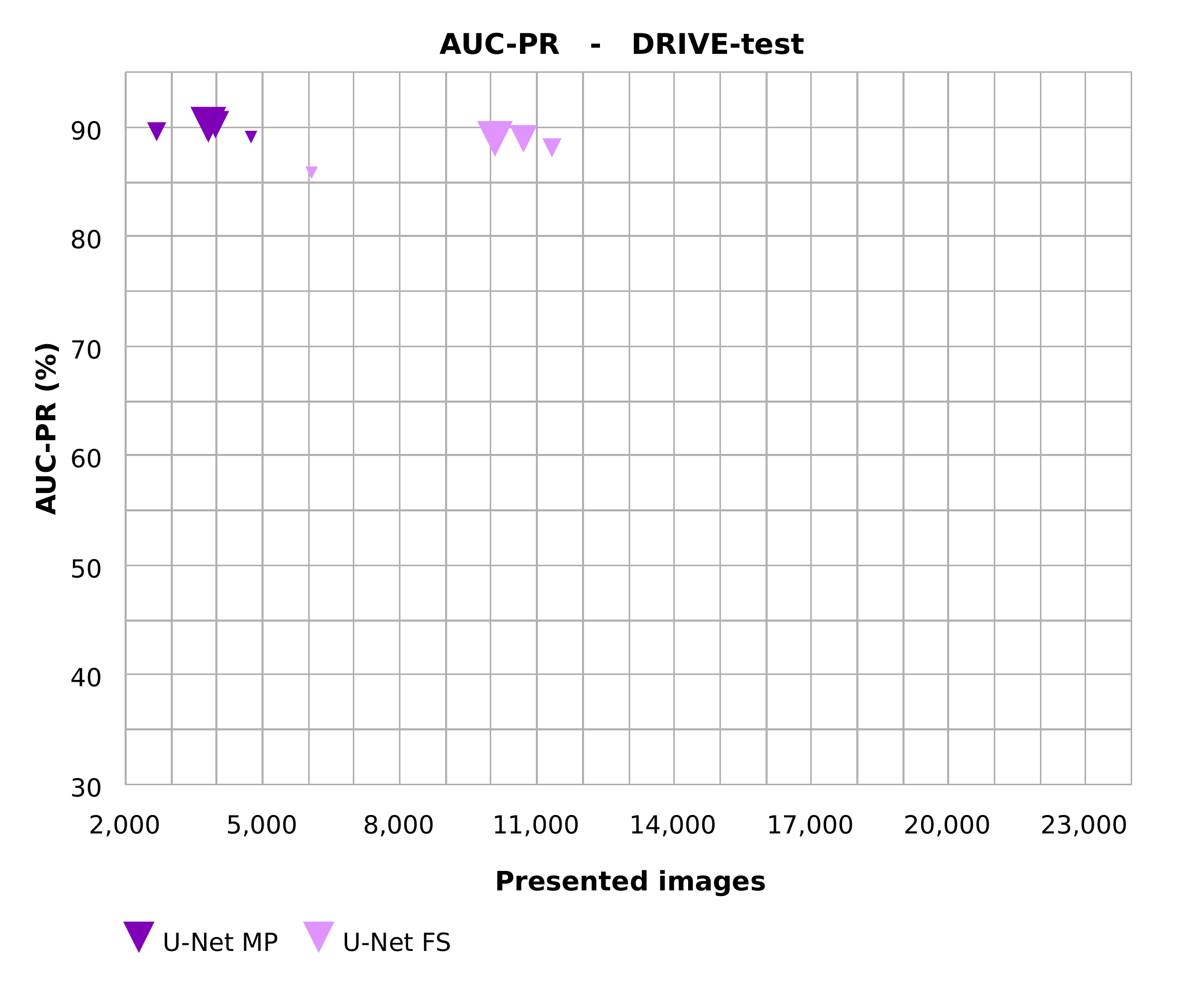}
  }
  \centerline{
  \includegraphics[width=0.33\textwidth, valign=t]{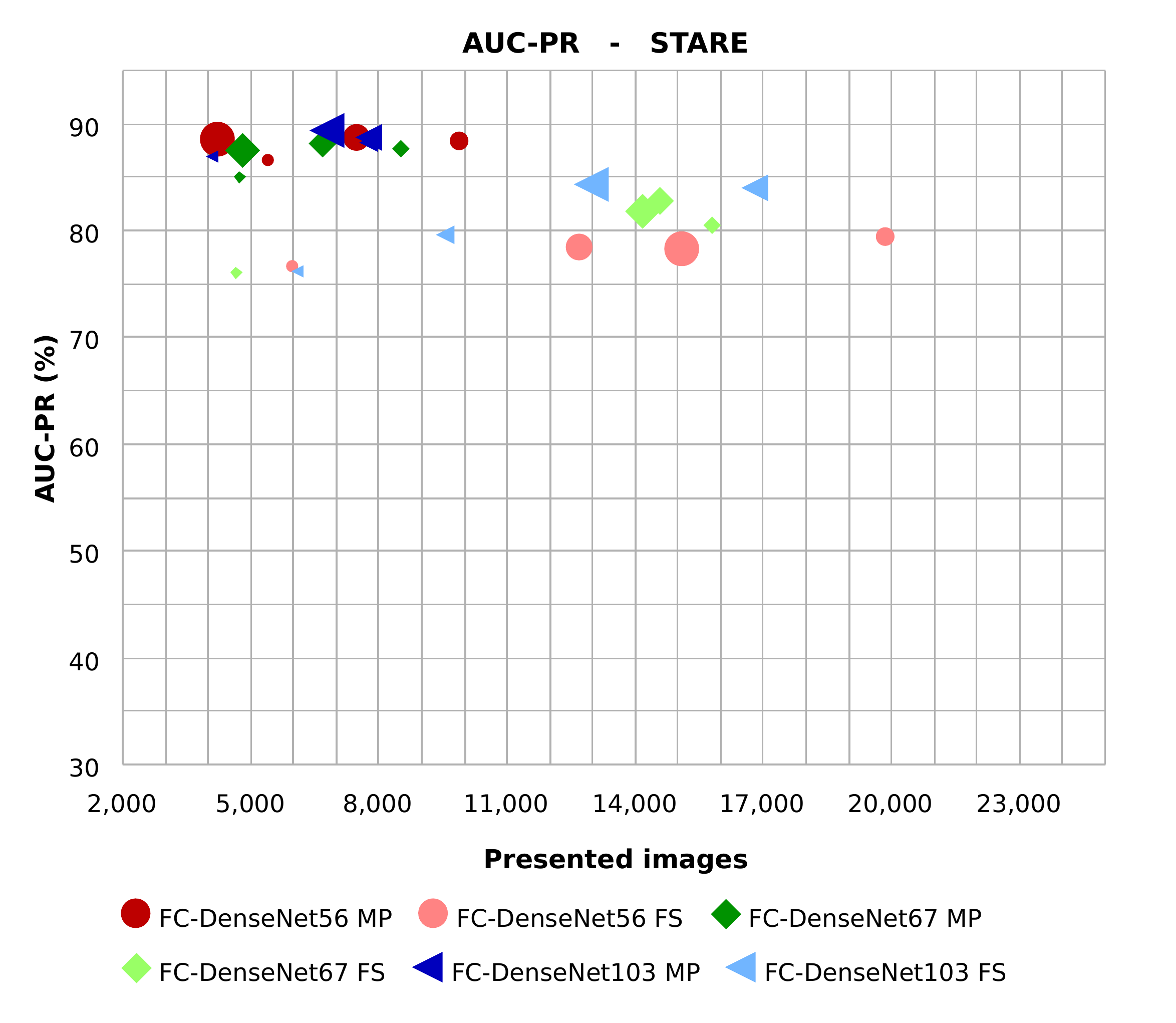}
  \includegraphics[width=0.33\textwidth, valign=t]{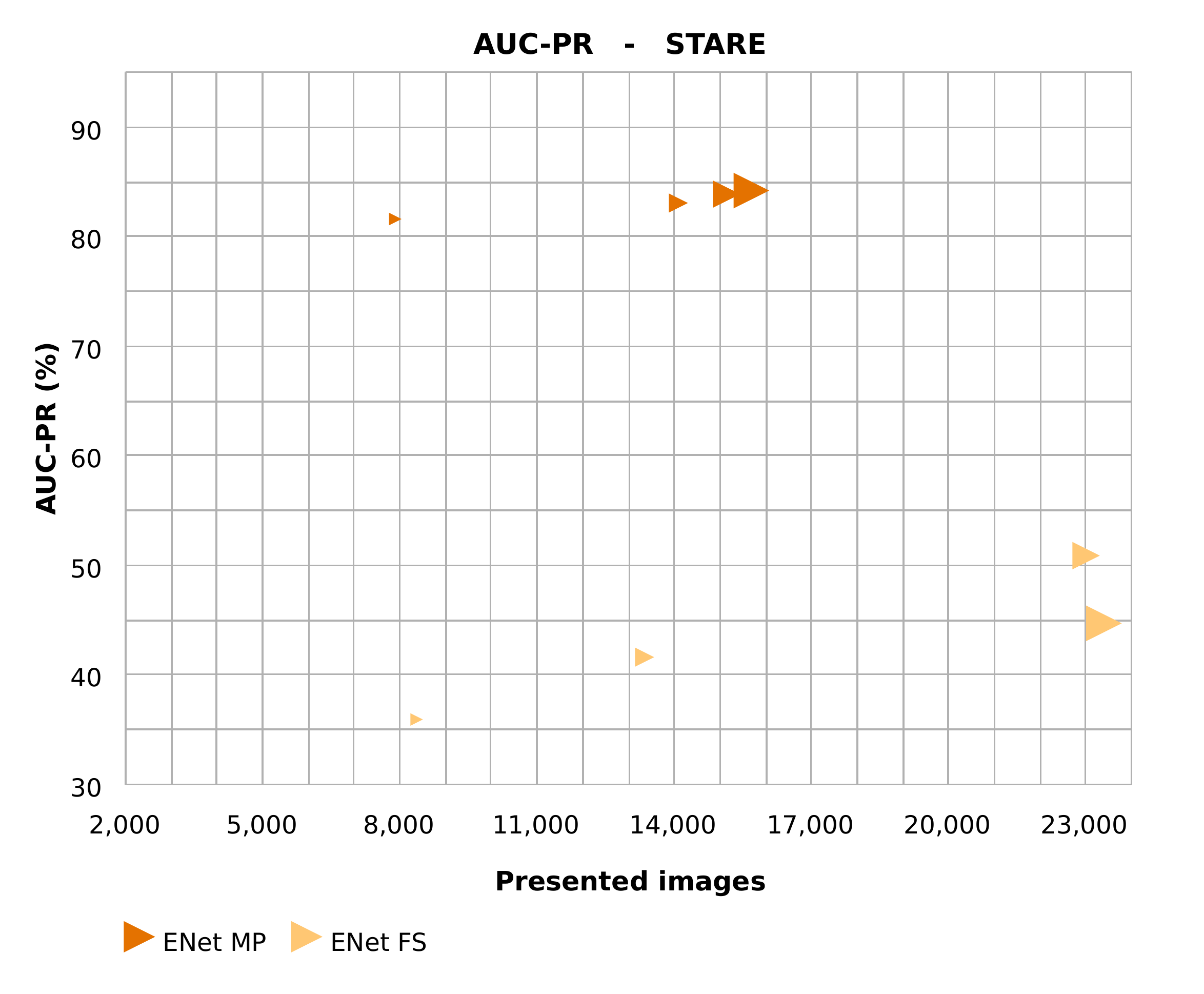}
  \includegraphics[width=0.33\textwidth, valign=t]{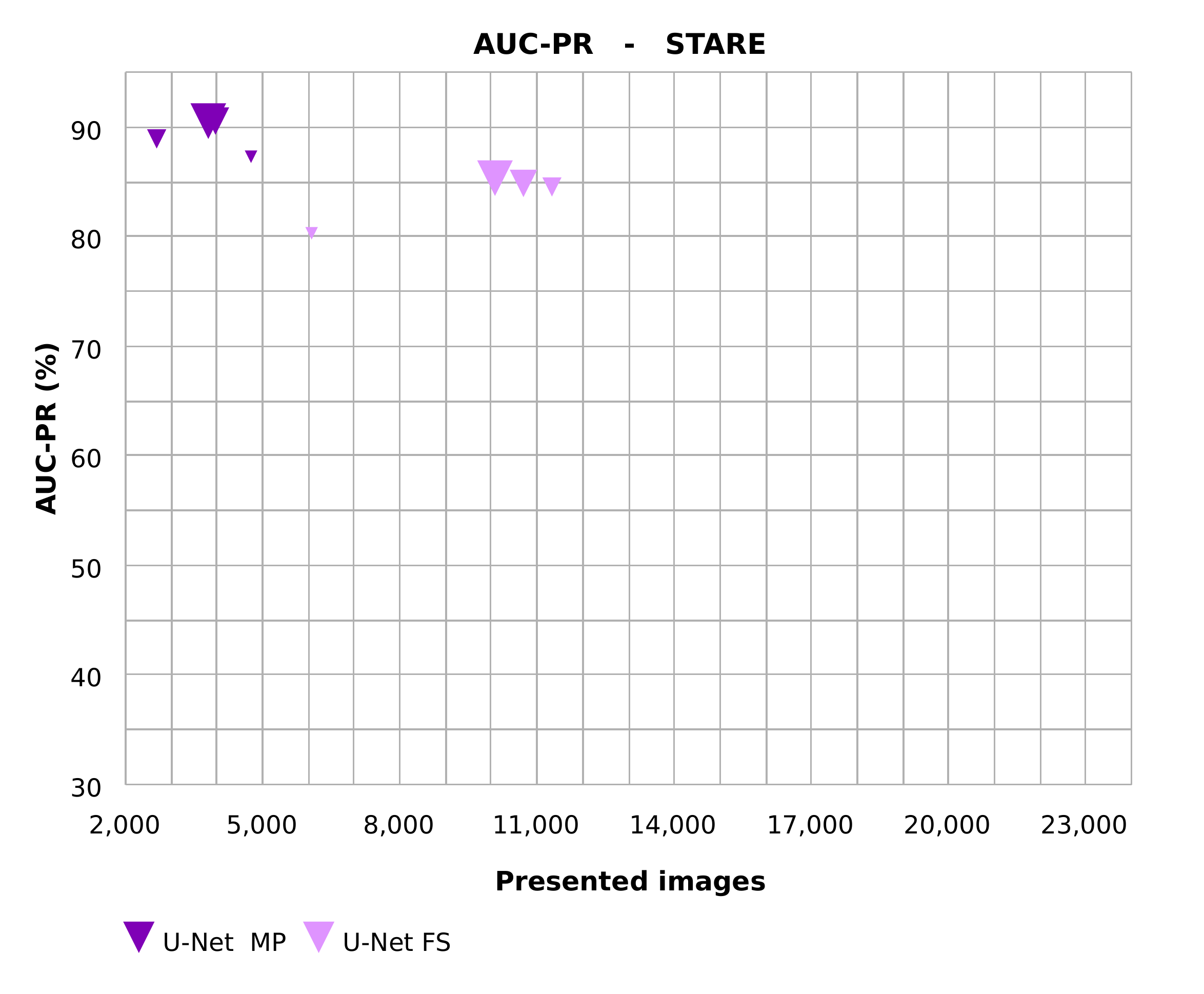}
  }
  \caption{AUC-PR values obtained by each trained model of the different architectures in DRIVE-test and STARE datasets. In the plots, this values are depicted against the number of images presented to the network during the training. The sizes of the bullets indicate the sizes of the training set used. Bright colors correspond to networks trained from scratch and dark colors correspond to networks trained following the proposed MP strategy.}
  \label{fig:2_PR_all_separated}
\end{figure*}

\begin{figure*}
  \centerline{\includegraphics[width=0.49\textwidth]{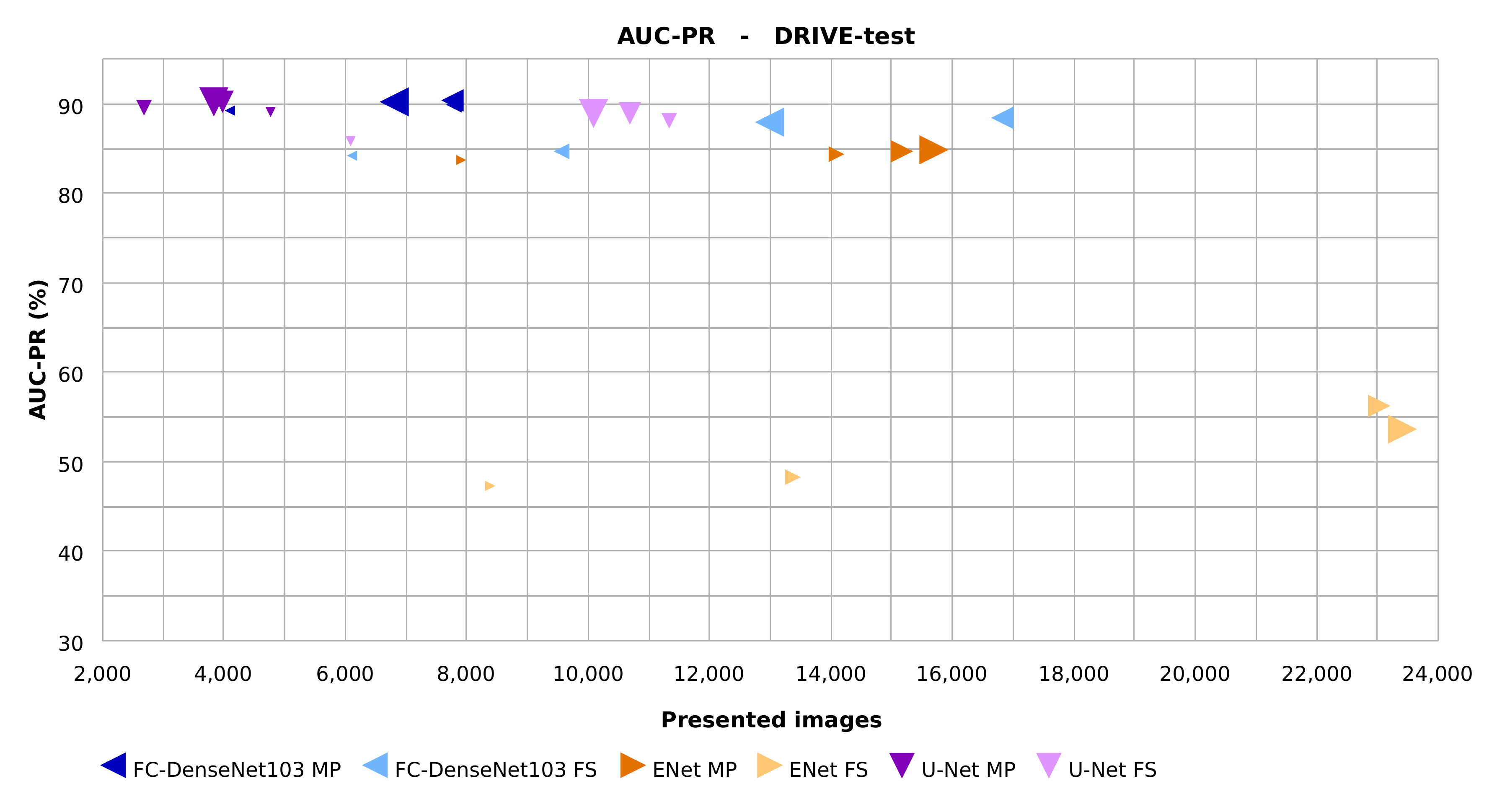}
  \includegraphics[width=0.49\textwidth]{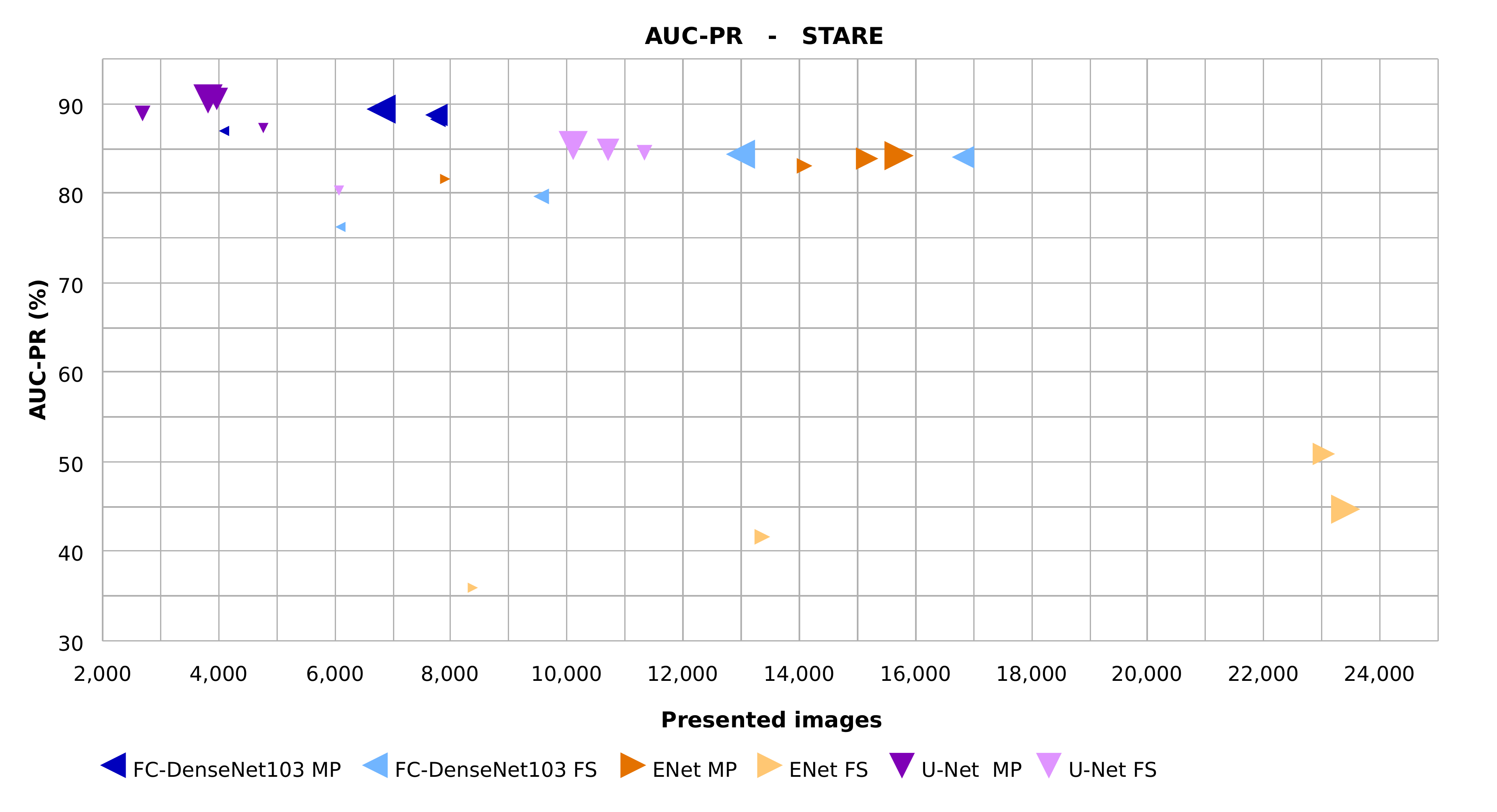}}
  \caption{AUC-PR values obtained by the U-Net, FC-DenseNet103 and ENet models merged for DRIVE-test and STARE datasets. In the plots, this values are depicted against the number of images presented to the network during the training. The sizes of the bullets indicate the sizes of the training set used. Bright colors correspond to networks trained from scratch and dark colors correspond to networks trained following the proposed MP strategy.}
  \label{fig:2_PR_all}
\end{figure*}

Figure \ref{fig:2_PR_all_separated} shows the AUC-PR values obtained by each trained model in DRIVE-test and STARE datasets. In the plots, these values are depicted against the number of images that were presented to the network during the training stage. Moreover, the sizes of the bullets indicate the sizes of the training sets. Complementarily, to allow the comparison between the different network architectures, Figure \ref{fig:2_PR_all} presents the results depicted in Figure \ref{fig:2_PR_all_separated} merged into two plots, one for DRIVE-test and other for STARE. For readability, only the 103 layer-version of FC-DenseNet was included. 

\begin{figure*}
  \centering
  \subfloat[]{\includegraphics[height=0.225\textwidth]{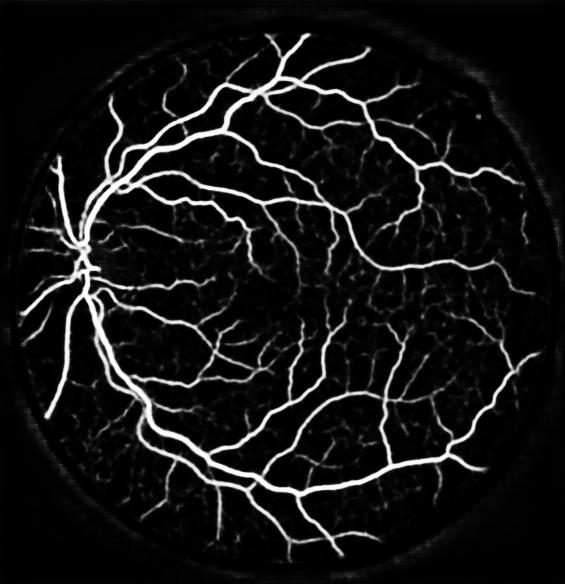}}
  \hspace{0.05cm}
  \subfloat[]{\includegraphics[height=0.225\textwidth]{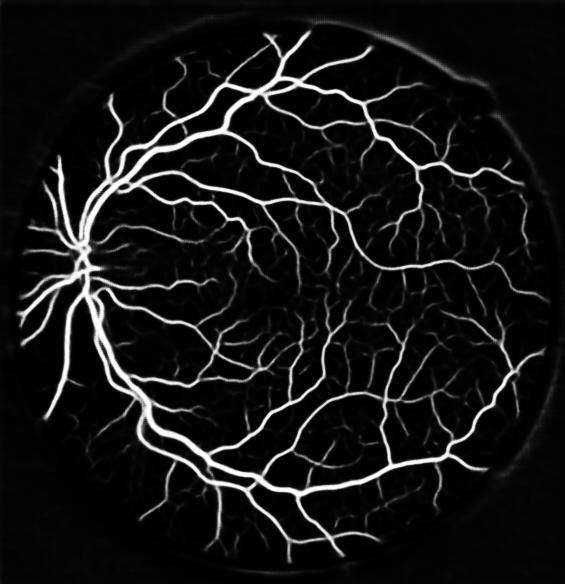}}
  \hspace{0.45cm}
  \subfloat[]{\includegraphics[height=0.225\textwidth]{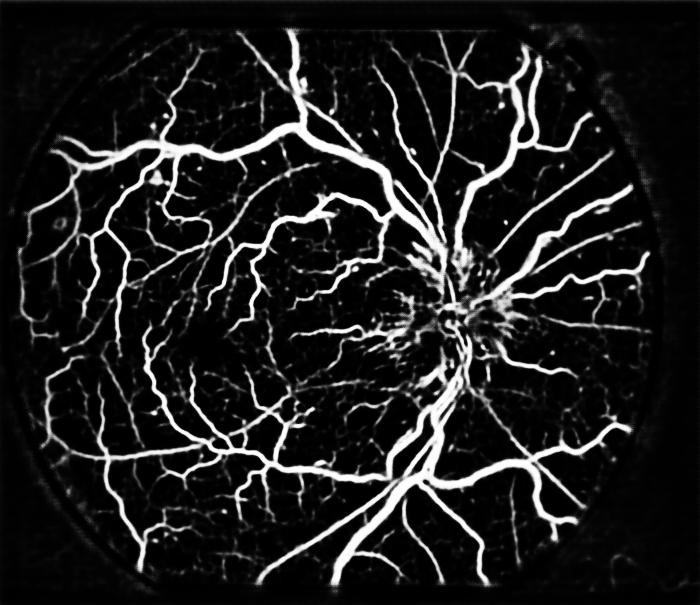}} 
  \hspace{0.05cm}
  \subfloat[]{\includegraphics[height=0.225\textwidth]{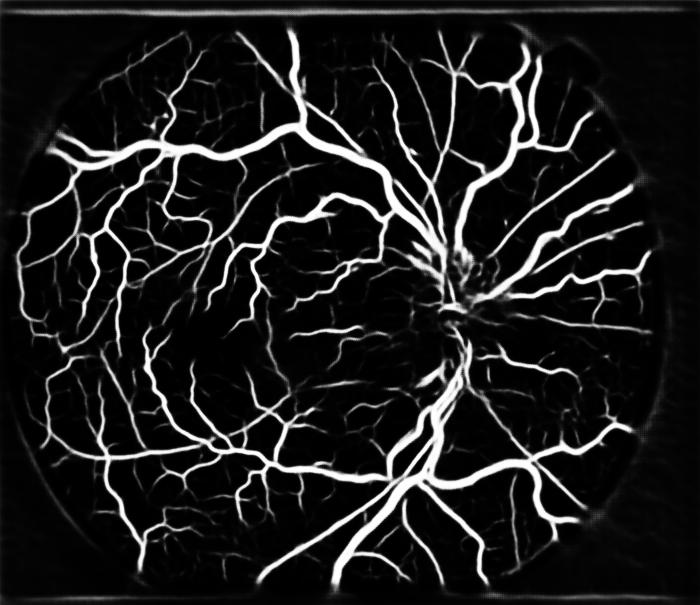}}

  \subfloat[]{\includegraphics[height=0.225\textwidth]{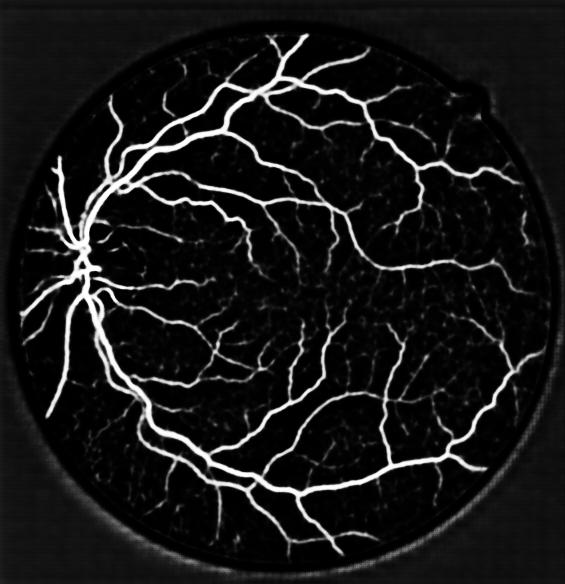}}
  \hspace{0.05cm}
  \subfloat[]{\includegraphics[height=0.225\textwidth]{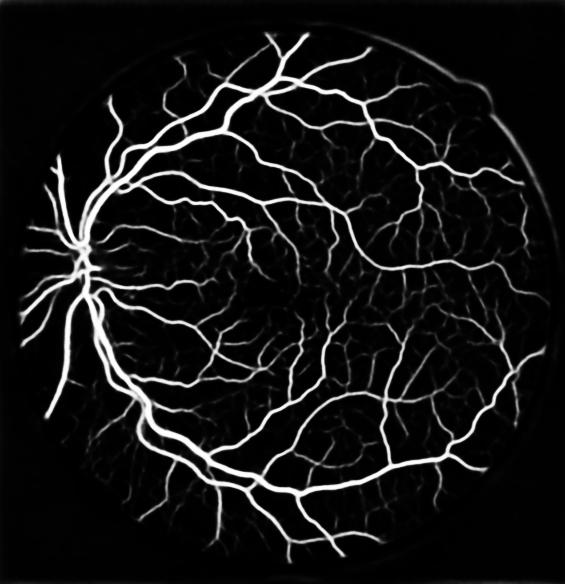}}
  \hspace{0.45cm}
  \subfloat[]{\includegraphics[height=0.225\textwidth]{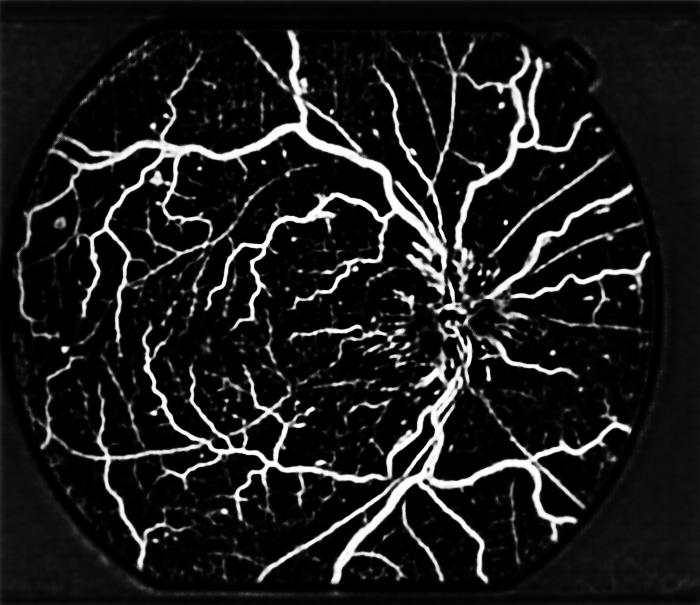}} 
  \hspace{0.05cm}
  \subfloat[]{\includegraphics[height=0.225\textwidth]{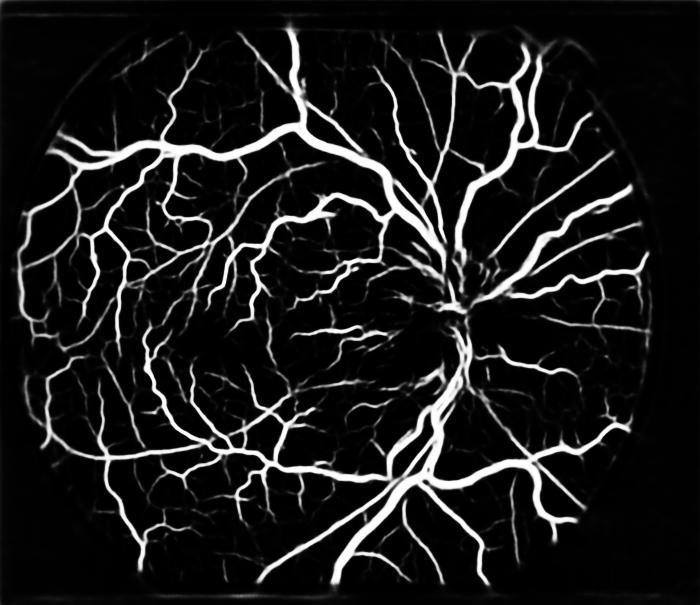}}

  \subfloat[]{\includegraphics[height=0.225\textwidth]{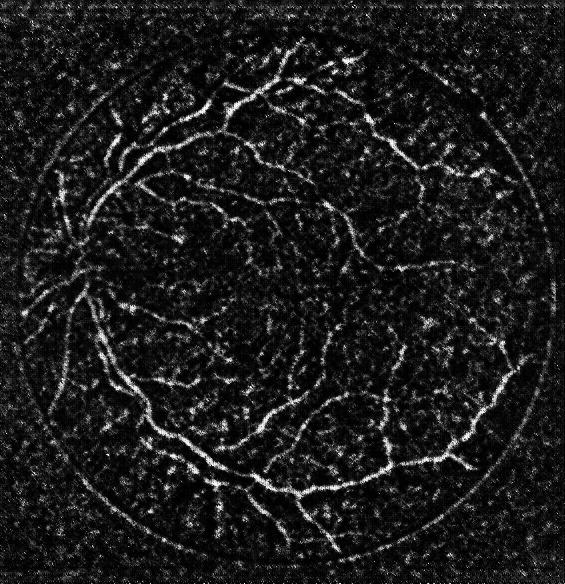}}
  \hspace{0.05cm}
  \subfloat[]{\includegraphics[height=0.225\textwidth]{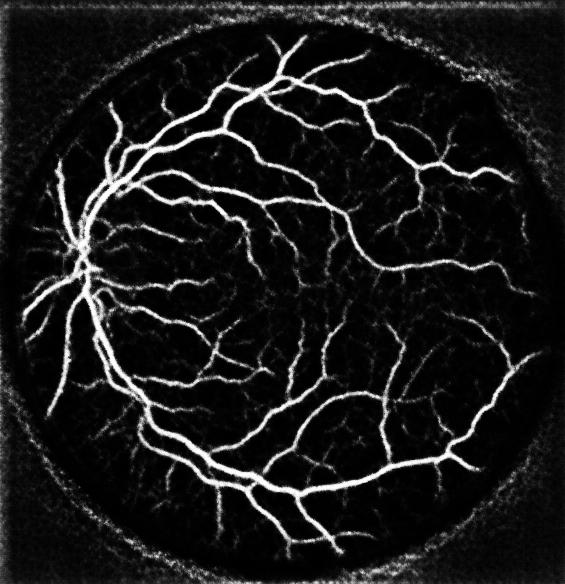}}
  \hspace{0.45cm}
  \subfloat[]{\includegraphics[height=0.225\textwidth]{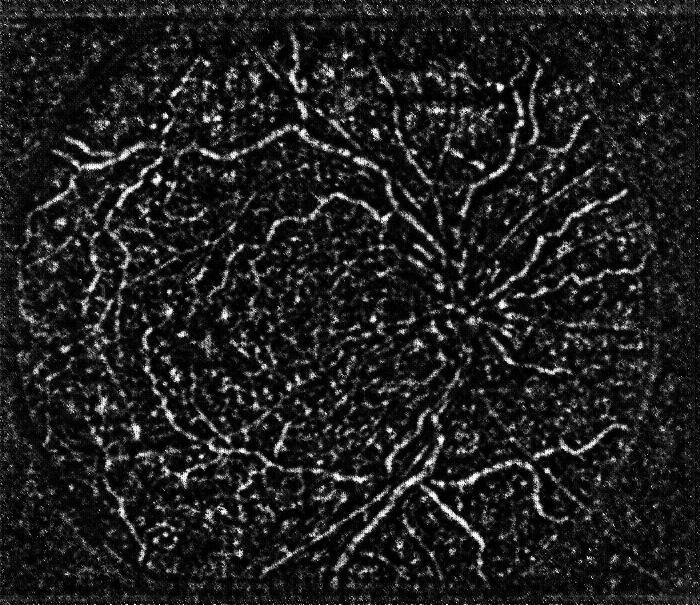}} 
  \hspace{0.05cm}
  \subfloat[]{\includegraphics[height=0.225\textwidth]{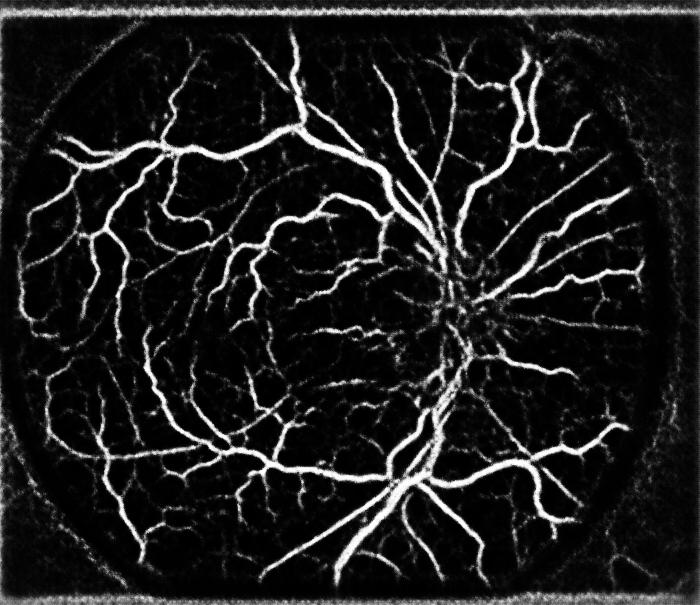}}
  
  \caption{Example of predicted segmentation masks for the test retinographies from STARE and DRIVE of Figure \ref{fig:examples_images_DRIVE-STARE}, using the compared architectures with and without pretraining (MP and baseline FS). In all cases, the networks were trained with a dataset of 15 images. (\nth{1} row) U-Net. (\nth{2} row) FC-DenseNet103. (\nth{3} row) ENet. (\nth{1} column) Results for the DRIVE retinography using the FS models. (\nth{2} column) Results for the DRIVE retinography using the MP models. (\nth{3} column) Results for the STARE retinography using the FS models. (\nth{4} column) Results for the STARE retinography using the MP models.}
  \label{fig:examples_all}
\end{figure*}

\begin{figure*}
  \centering
  \subfloat[]{\includegraphics[height=0.225\textwidth]{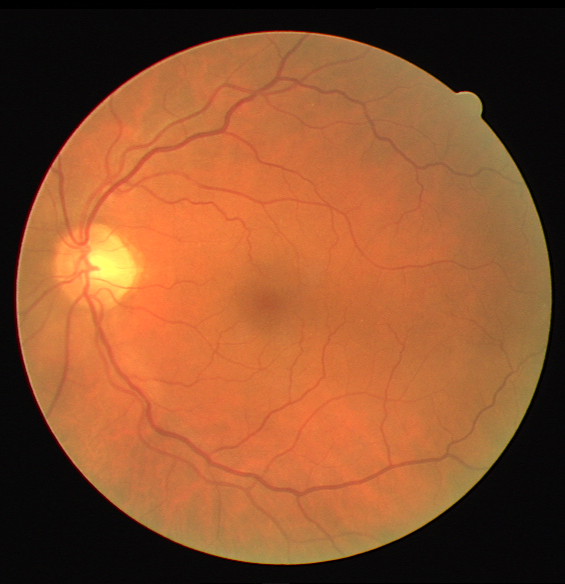}}
  \hspace{0.05cm}
  \subfloat[]{\includegraphics[height=0.225\textwidth]{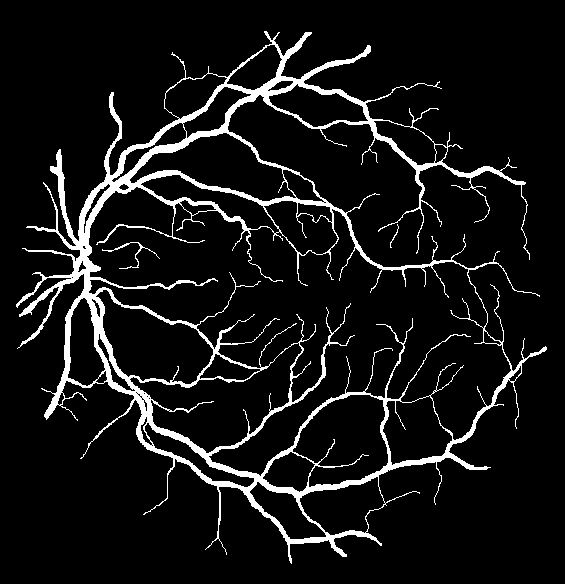}}
  \hspace{0.45cm}
  \subfloat[]{\includegraphics[height=0.225\textwidth]{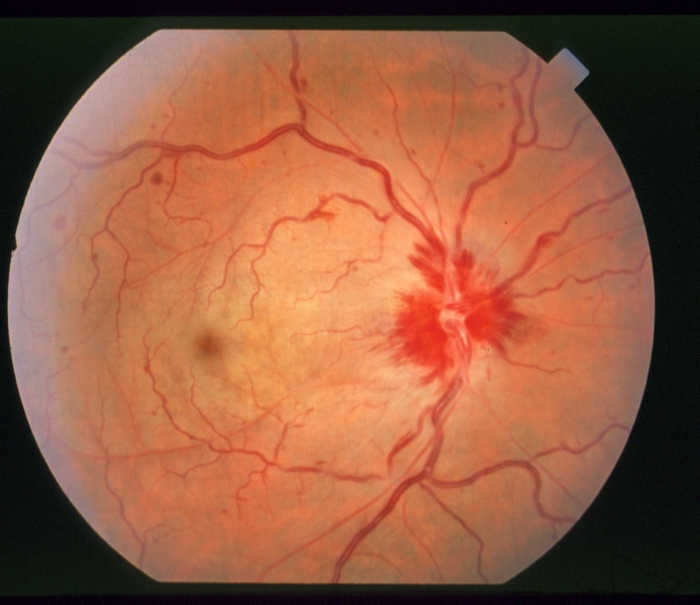}} 
  \hspace{0.05cm}
  \subfloat[]{\includegraphics[height=0.225\textwidth]{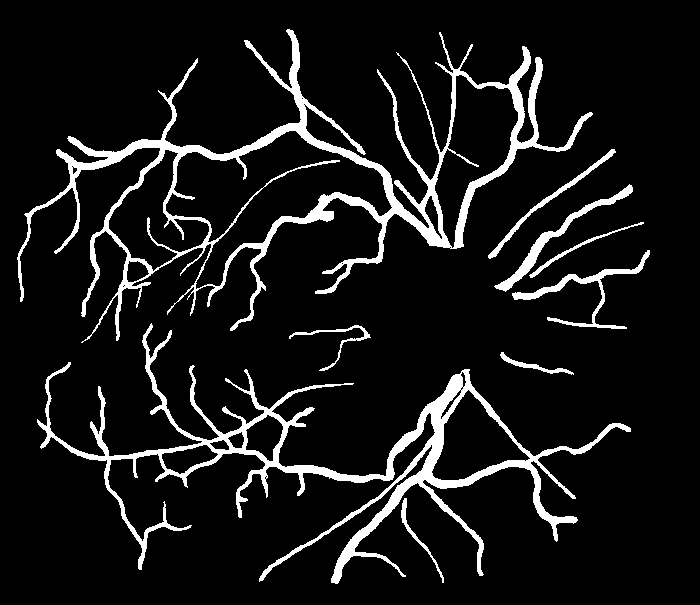}}

  \caption{Example test set retinographies from STARE and DRIVE with their respective ground truth vascular segmentation masks. (a) Example retinography from DRIVE. (b) Ground truth for the DRIVE retinography. (c) Example retinography from STARE. (d) Ground truth for the STARE retinography.}
  \label{fig:examples_images_DRIVE-STARE}
\end{figure*}

Also, Figure \ref{fig:examples_all} presents representative examples of segmentation masks obtained by the analysed networks trained from scratch (columns 1 and 3) and with the proposed MP approach (columns 2 and 4) in the couple of illustrative retinographies from the DRIVE-test and STARE datasets shown in Figure \ref{fig:examples_images_DRIVE-STARE}. In all the cases, the size of the training set was 15 images. From FC-DenseNet models, we have chosen FC-DenseNet103, as it provided the best results.

Generally, as observed in the Figures, all the architectures greatly benefit from using pretrainings from the multimodal reconstruction task, independently of the dataset that was used during the evaluation. In all the cases, resulting AUC-PR is higher and the required training time of the networks is remarkably shorter (see Figure \ref{fig:2_PR_all_separated}). Moreover, the provided examples (Figure \ref{fig:examples_all}) clearly show that the quality of the vascular extractions is higher when the pretraining models are used. In this sense, one of the main improvements of the MP approach with respect to the FS baseline is the consistency in the vessel continuity, as it preserves significantly better continuous vascular trees. Additionally, the networks that were trained following the MP approach are more sensitive to the small and thin vasculature, considered the main complexity of this issue. Another important complexity is the presence of other structures, as those produced by pathological scenarios (\textit{e.g.} drusen, microaneurysms or haemorrhages). These challenging scenarios are better handled by the networks trained with the proposed multimodal transfer learning approaches. This improvement of the performance is specially intense in the case of the ENet, a network with lower capacity, where the multimodal pretraining is crucial to reach any acceptable result.

Although Figure \ref{fig:examples_all} only shows an example for models trained with 15 images, the stated differences between the segmentation masks obtained by MP and FS models are the same for all the studied cases. These differences, besides, are more remarkable when the networks are trained with a training set of size 1, as can be deduced from the results depicted in Figure \ref{fig:2_PR_all_separated}.

Another relevant point that can be observed in the plots is the robustness of the networks to different sizes of the training set. The plots of Figures \ref{fig:2_PR_all_separated} and \ref{fig:2_PR_all} show that the proposed MP models are, again, significantly better than the baseline FS. When pretraining is used, the results of the same network trained with different dataset sizes oscillate much less that the same network trained from scratch.

In addition to the previous discussion, it is also interesting to notice how FC-DenseNet benefits from its greater capacity. In Figure \ref{fig:2_PR_all_separated}, it can be observed that, in most of the cases, the FC-DenseNet103 results are better than those of FC-DenseNet56 and FC-DenseNet67.

Complementarily, as said, the plots depicted in Figure \ref{fig:2_PR_all} show the results of the different architectures all together. These plots show clearly that the results of the ENet are significantly far from those of the U-Net and the FC-DenseNet, given its limited capacity. Despite this, we would like to remark that the provided results are still satisfactory, considering that it represents an architecture with a more reduced computational requirements. Moreover, we can see that the results of the U-Net and the FC-DenseNet models are very similar for DRIVE and a little more different for STARE in favor of U-Net. In this case, the higher difficulty of STARE seems to affect the FC-DenseNet more than the U-Net. Thus, it can be interpreted that the U-Net presents a greater capacity of generalization, being more robust to the presence of severe signs of pathologies and injuries in the original retinographies (in DRIVE, there are also retinographies with signs of diabetic retinopathy, but they are not so remarkable). Additionally, U-Net needs fewer epochs to converge to valid solutions.

Based on all those facts, we consider the U-Net architecture the most suitable for being pretrained in this multimodal context and refined in the task of the retinal vascularity segmentation, presenting an adequate balance of dimensionality, capacity of generalization and performance in the issue. However, we have to consider the larger parameter set that needs to be handled. The FC-DenseNet, with less parameters, also provided a competitive performance but without taking any representative advantage of its characteristic potential and dimensionality. Finally, as expected, the ENet models offered the worse results. However, the obtained performance is still satisfactory in the vascular extraction task when the pretraining is used. Thus, this network architecture could be an interesting option, specially when the computational requirements would be a limitation.


\section{CONCLUSIONS}

In this work, we presented and studied multimodal transfer learning-based approaches for retinal vascular segmentation. These approaches benefit from a pretraining in a multimodal reconstruction task that uses, as reference, two classical and widely used eye fundus image modalities: retinographies and angiographies. This latter retinal modality is specially useful for the retinal vessel analysis, as it specially highlights the vascular structure. To measure the potential and capabilities of the proposed approaches, we also trained from scratch the corresponding models used in the proposed approaches.

The results that were obtained by the different approaches showed that the use of pretrained models in the multimodal reconstruction task has proven to be adequate independently of the network architecture, specially in comparison to those trained from scratch, providing better results with less required training times on the target task. In addition, the experimentation with various dataset sizes allowed us to notice the great robustness against small training datasets of the networks with transferred learning. Thus, the use of pretrained models in the multimodal reconstruction task emerges as a valuable option with scarcity of annotated data in the target task. As a self-supervised approach, multimodal reconstruction does not depend on experts to annotate the image datasets. In this way, we can use a significant number of pairs retinography-angiography without any annotation (relatively easy to obtain) in the multimodal reconstruction task, and the few available annotated data in the target task. Thus, the network performance is highly improved without needing extra participation of experts. This is one of the most important results.

Complementarily, based on the results of the different tested models, U-Net was the one that provided the best performance, despite the larger parameter set, providing an adequate balance of dimensionality and capacity of generalization, also in less training epochs. The tested FC-DenseNet configurations, with less parameters, also provided a competitive performance but without taking advantage of its significant capacity. In the case of the ENet, the performance of this architecture was worse, as expected, but thanks to multimodal pretraining, it was still satisfactory considering its lower capacity. Thus, this architecture could be an interesting option when the computational requirements would be a limitation.


\ack{This work is supported by Instituto de Salud Carlos III, Government of Spain, and the European Regional Development Fund (ERDF) of the European Union (EU) through the \mbox{DTS18/00136} research project, and by Ministerio de Ciencia, Innovaci\'on y Universidades, Government of Spain, through the \mbox{DPI2015-69948-R} and \mbox{RTI2018-095894-B-I00} research projects. The authors of this work also receive financial support from the ERDF and European Social Fund (ESF) of the EU, and Xunta de Galicia through Grupo de Referencia Competitiva, ref. \mbox{ED431C 2016-047}, and the predoctoral grant contract ref. \mbox{ED481A-2017/328}.}


\clearpage

\bibliography{ecai}
\end{document}